\def\.{\mbox{ \tiny{ $^\bullet$} }}

\def\eps{\epsilon}
\def\ko{k_0}
\def\lambdao{\lambda_0}
\def\etao{\eta_0}

\documentclass[12pt]{iopart}
\usepackage{epsfig}

\begin{document}
\title[The negative index of refraction demystified]{The negative
index of refraction demystified}

\author{Martin~W~McCall\dag\footnote[4]{To
whom correspondence should be addressed (m.mccall@ic.ac.uk)},
Akhlesh~Lakhtakia\ddag\ and Werner~S~Weiglhofer\S }

\address{\dag\ Department of Physics, The Blackett Laboratory,
Imperial College of Science, Technology and Medicine, Prince
Consort Road, London SW7 2BZ, Great Britain}

\address{\ddag\ CATMAS~---~Computational \& Theoretical Materials Sciences
Group, Department of Engineering Science and Mechanics,
Pennsylvania State University, University Park, PA 16802--6812,
USA}

\address{\S\ Department of Mathematics, University of Glasgow,
University Gardens, Glasgow G12 8QW, Great Britain}

\begin{abstract}
We study electromagnetic wave propagation in mediums in which the
effective relative permittivity and the effective relative
permeability are allowed to take any value in the upper half of
the complex plane. A general condition is derived for the phase
velocity to be oppositely directed to the power flow.  That
extends the recently studied case of propagation in mediums for
which the relative permittivity and relative permeability are both
simultaneously negative, to include dissipation as well. An
illustrative case study demonstrates that in general the spectrum
divides into five distinct regions.
\end{abstract}


\vskip 1.0cm Accepted for: {\EJP}

\maketitle

\section{Introduction}

Materials with either negative real relative permittivity or
negative real relative permeability at a certain frequency are not
uncommon. Consideration of materials for which both these
quantities are simultaneously negative and real--valued, commenced
with Veselago's paper of 1968 \cite{Veselago}.  Although he
pointed out many unusual properties of such materials, including
inverse refraction, negative radiation pressure, inverse Doppler
effect, over three decades ago, the considerations were completely
speculative in view of the lack of a material, or even a
nonhomogeneous composite medium, with a relative permittivity
having a negative real part and a very small imaginary part. A
breakthrough was achieved by Smith {\em et al.\/} \cite{Schultz1},
who, developing some earlier ideas by Pendry {\em et al.\/}
\cite{Pendry1, Pendry2, Pendry5}, presented evidence for a weakly
dissipative composite medium displaying negative values for the
real parts of its {\em effective\/} permittivity and {\em
effective\/} permeability. Their so--called {\em meta}--material
consists of various inclusions of conducting rings and wires
embedded within printed circuit boards. Their conclusions were
based on observations from three separate composite mediums.
\begin{itemize}
\item\underline{Medium 1}
consisted of a lattice of ring--like inclusions, which for a
certain field configuration was presumed to have a resonant
relative permeability ${\mu}_{eff}(\omega )$ of the form
\cite{Schultz1, Pendry5}
\begin{equation}
\label{mu} {\mu}_{eff}(\omega) = 1 -
\frac{F{\omega}^2}{{\omega}^2-{\omega}_0^2+i\omega \Gamma}~,
\end{equation}
where the resonance frequency ${\omega}_0$ depends principally on
the geometry of the rings. In this model, dissipation is
facilitated by $\Gamma$, and $F$ ($0<F<1$) is the ratio of the
area occupied by a ring and that of a unit cell. For weak
dissipation, the real part of ${\mu}_{eff}$ is negative for
${\omega}_0 < \omega < {\omega}_0 /\sqrt{1-F}$.
\item\underline{Medium 2}
consisted of an included matrix of wires. The effective relative
permittivity ${\eps}_{eff}$ of this composite medium supposedly
displays plasma--like behaviour according to
\begin{equation}
\label{eps} {\eps}_{eff} (\omega ) = 1 -
\frac{{\omega}_p^2}{{\omega}^2}~,
\end{equation}
the effective plasma frequency ${\omega}_p$ being dependent on the
geometry. In such a medium, ${\eps}_{eff}$ is negative for $\omega
< {\omega}_p$.
\item\underline{Medium 3}
combined the first two, it being postulated that the combination
would exhibit negative real--valued permeability and negative
real--valued permittivity within a certain frequency range.
Although some numerical and experimental evidence was presented in
2000 \cite{Schultz1}, the most striking report appeared the
following year \cite{Schultz2} and gave preliminary indications of
the observation of the so--called negative index of refraction.
\end{itemize}
Other types of nanostructural combinations with similar response
properties can also be devised \cite{dewar}.

The emergence of a clear terminology is often a difficult process
with regards to scientific findings relating to novel effects,
something that is also apparent in the present instance. The
original classification of the materials exhibiting the effects
discussed labelled them {\em left--handed electromagnetic
materials\/} \cite{Veselago}. But chiral materials are important
subjects of electromagnetics research and the terms {\em
left--handedness\/} and {\em right--handedness\/} have been
applied to the molecular structure
   of such materials for well over a century \cite{SPIE}.
The continued use of the term {\em left--handed materials\/}
(LHMs) for achiral materials in, for example, \cite{Schultz1,
Schultz2, Smith} will thus confuse the crucial issues. Other
authors \cite{Lindell} are promoting the term {\em backward
medium\/} (BW) which presumes  the {\em a priori\/} definitions of
forward and backward directions. In the most recent contribution
\cite{Ziol} (which also provides the most extensive theoretical
and numerical analysis of the negative index of refraction to
date), the authors introduce the technical term {\em double
negative\/} (DNG) medium to indicate that the real parts of both
permittivity and permeability are negative. While sensible enough,
such nomenclature conceals the importance of dissipative effects.

In time, a consensus about terminology will undoubtedly emerge;
and it is not our aim to contribute to this particular discussion.
Instead the purpose of this note is pedagogical. In the first
instance, it is important that dissipation be included in the
analysis. This was largely neglected in the reports cited earlier,
with the exemption of the most recent study  \cite{Ziol}.

Secondly, it is desirable to derive the general condition for the
type of {\em anomalous\/} propagation that is characteristic of
the considered materials: namely, where the phase velocity is
directed oppositely to the power flow.

\section{Plane Wave Propagation}

Consider a plane wave propagating along the $z$ axis in a linear,
homogeneous, isotropic, dielectric--magnetic medium whose relative
permittivity and relative permeability are denoted by $\epsilon_r$
and $\mu_r$, respectively. An $\exp(-i\omega t)$ time--dependence
is assumed here. Then
\begin{eqnarray}
&&{\bf E}(z) = A\, \exp (i\ko n z) \, {\bf u}_x \,,
\\
&&{\bf B}(z) =\frac{1}{i\omega}\, \nabla\times {\bf E}(z)
=\frac{n\ko}{\omega}\,A\, \exp (i\ko n z) \, {\bf u}_y \,,
\\
&&{\bf H}(z) = \frac{n}{\mu_r\etao} \, A\, \exp (i\ko n z) \, {\bf
u}_y \,,
\end{eqnarray}
where $\ko$ is the free--space wavenumber, $\etao$ is the
intrinsic impedance of free space, and $n^2 = \eps_r\mu_r$.
Consequently, the Poynting vector is parallel to the $z$ axis and
its time--average is given as
\begin{equation}
\label{power} P_z(n) = \frac{1}{2}\,{\bf u}_z\. {\rm Re} \Bigl[
{\bf E}(z)\times {\bf H}^\ast (z)\Bigr] = {\rm Re} \left[
\frac{n}{\mu_r}\right]\, \frac{\vert
A\vert^2}{2\etao}\,\exp(-2\ko{\rm Im}[n]z)\,~,
\end{equation}
where ${\rm Re}[\cdot]$ and ${\rm Im}[\cdot]$, respectively,
denote the operations of taking the real and the imaginary part,
whilst $^\ast$ indicates complex conjugation.

Let us now assume a Lorentzian model for $\eps_r$ and $\mu_r$.
This will  include the specified forms (\ref{mu}) and (\ref{eps})
as special cases.  Dissipation results from the imaginary parts of
$\eps_r$ and $\mu_r$ whilst causality dictates that ${\rm
Im}[\mu_r]
>0$ and ${\rm Im}[\eps_r] >0$, so that $\eps_r$ and $\mu_r$ lie in
the upper half of the complex plane.

However, there are two resultant complex refractive indexes,
$n_{\pm} = \pm \sqrt{\eps_r \mu_r}$, of which $n_+$ lies in the
upper half of the complex plane and $n_-$ in the lower half. The
situation is summarized in Figure 1 for which the resonant form of
(\ref{mu}) was used as representative of both $\eps_r (\omega )$
and $\mu_r (\omega )$. Of course, the resonances of $\eps_r
(\omega )$ and $\mu_r (\omega )$ are unlikely to coincide, so that
for a particular value of $\omega$, the arrows corresponding to
$\eps_r$ and $\mu_r$ will not be necessarily parallel. Only the
upper half of the complex plane is shown in the figure.

Now $n_\pm$ may be written as
\begin{equation}
n_\pm = \pm n_0 \exp {i{\phi}_n}\,,
\end{equation}
where
\begin{equation}
n_0 = + \sqrt{|\eps_r ||\mu_r |} \, , \qquad {\phi}_n =
\frac{{\phi}_{\eps}+{\phi}_{\mu}}{2} \, .
\end{equation}
Here ${\phi}_{\eps}$ and ${\phi}_{\mu}$, representing the
arguments of $\eps_r$ and $\mu_r$ respectively, must obey the
conditions $0 \leq {\phi}_{\eps,\mu} \leq \pi$. Consequently, $0
\leq {\phi}_n \leq \pi$. We then always have
\begin{equation}
{\rm Re}\left[\frac{n_+}{\mu_r}\right] > 0 \quad {\rm i.e.}\
P_z(n_+) >0
\end{equation}
and also
\begin{equation}
{\rm Re}\left[\frac{n_-}{\mu_r}\right]< 0 \quad {\rm i.e.}\
P_z(n_-) <0 \,.
\end{equation}
Thus the choice $n_+$ {\em always\/} relates to power flow in the
$+z$ direction, whilst $n_-$ {\em always\/} relates to power flow
in the $-z$ direction. Since necessarily ${\rm Im}[n_+] > 0$ and
${\rm Im}[n_-]< 0$, power flow is always in the direction of
exponential decrease of the fields' amplitudes.

We can now identify when the phase velocity is opposite to the
direction of power flow. This occurs whenever ${\rm Re}[n_+] < 0$
(and consequently ${\rm Re}[n_-]  > 0$, also). After setting
\begin{equation}
\eps_r = \eps'_r+i\eps''_r \, , \qquad \mu_r = \mu'_r+i\mu''_r \,
,
\end{equation}
(where $\eps'_r, \eps''_r$ and $\mu'_r, \mu''_r$ are the real and
imaginary parts of the relative permittivity and the relative
permeability, respectively), the following condition is
straightforwardly derived for such propagation:
\begin{equation}
\label{inequality} \left [ +{\left( {\eps'_r}^2+{\eps''_r}^2
\right )}^{1/2} - {\eps'_r} \right ] \left [ +{\left( {\mu'_r}^2
+{\mu''_r}^2   \right )}^{1/2}- {\mu'_r} \right ] >
\eps''_r\mu''_r~~.
\end{equation}

Before turning to a fully illustrative example in the proceeding
section, let us investigate some immediate repercussions of the
inequality (\ref{inequality}) which is central to this paper.
\begin{itemize}

\item
Consider, in the first instance, the behaviour at a resonance of
the relative permittivity, i.e. $\eps'_r= 0$, $\eps''_r > 0$.
Then, (\ref{inequality}) reduces to
\begin{equation}
 \left [ +{\left( {\mu'_r}^2
+{\mu''_r}^2   \right )}^{1/2}- {\mu'_r} \right ] > \mu''_r \, ,
\end{equation}
an inequality that is always fulfilled when $\mu'_r < 0$.

Likewise, at a resonance of the relative permeability, i.e.
$\mu'_r= 0$, $\mu''_r > 0$, (\ref{inequality}) is fulfilled
whenever $\eps'_r < 0$.

\item
Further insight into inequality (\ref{inequality}) can be gained
by requiring that
\begin{equation}
\label{inequalitye}
  +{\left( {\eps'_r}^2+{\eps''_r}^2 \right )}^{1/2} > {\eps'_r}
+\eps''_r
\end{equation}
and
\begin{equation}
\label{inequalitym}
  +{\left( {\mu'_r}^2+{\mu''_r}^2 \right )}^{1/2} > {\mu'_r}
+ \mu''_r
\end{equation}
simultaneously hold. Consequently, (\ref{inequality}) is
definitely satisfied. It should be remarked though, that the
parameter space of the permittivity and permeability that fulfils
(\ref{inequalitye}) and (\ref{inequalitym}) is only a subset of
the one fulfilling (\ref{inequality}). In any case,
(\ref{inequalitye}) holds if and only if $\eps'_r < 0$, and
(\ref{inequalitym}) holds if and only if $\mu'_r < 0$ (we remind
the reader that $\eps''_r > 0$, $\mu''_r> 0$ because of causality
requirements). We note that $\eps'_r < 0$ and $\mu'_r < 0$ can
only occur close to absorption resonances (as discussed in the
previous item).

\item
Finally, consider an electromagnetic wave propagating in a plasma
below the plasma frequency ($\eps'_r<0, \mu'_r=1$) and in which
dissipation is very small ($\eps''_r \ll 1$, $\mu''_r \ll 1$).
Straighforward Taylor expansions reduce inequality
(\ref{inequality}) to
\begin{equation}
\label{ineq}
 \vert \eps'_r \vert > \frac{\eps''_r}{\mu''_r} \, .
\end{equation}
Therefore, the existence of the type of anomalous propagation
being studied here depends in this case crucially on the ratio of
the imaginary parts of the relative permittivity and relative
permeability. Whether the criterion (\ref{ineq}) is satisfied or
not, the power flow in this case is in any case small.

\end{itemize}

\section{A detailed illustrative case study}

Let us exemplify the foregoing in detail by an explicit invocation
of the Lorentz model for both $\eps_r$ and $\mu_r$; thus,
\begin{eqnarray}
\label{Leps} \eps_r(\lambdao) = 1 + \frac{p_e}{1+\left(N_e^{-1} -
i\lambda_e\lambdao^{-1}\right)^2}\,,
\\
\label{Lmu} \mu_r(\lambdao) = 1 + \frac{p_m}{1+\left(N_m^{-1} -
i\lambda_m\lambdao^{-1}\right)^2}\,.
\end{eqnarray}
Here $\lambdao=2\pi/\ko$ is the free--space wavelength, $p_{e,m}$
are the oscillator strengths,
$\lambda_{e,m}(1+N_{e,m}^{-2})^{-1/2}$ are the resonance
wavelengths, while $\lambda_{e,m}/N_{e,m}$ are the resonance
linewidths.

Figures 2a and 2b comprise plots of the real and imaginary parts
of $\eps_r$ and $\mu_r$ as functions of $\lambdao$, when $p_e=1$,
$p_m=0.8$, $N_e = N_m=100$, $\lambda_e=0.3$~mm and
$\lambda_m=0.32$~mm. Clearly, five separate spectral regions can
be identified in Figure 2. At the either extremity of the
horizontal axis are the two regions wherein $\eps'_r >0$ and
$\mu'_r >0$. In the neighbourhood of $\lambdao=0.22$~mm, $\eps'_r
<0$ but $\mu'_r
>0$. Both $\eps'_r <0$ and  $\mu'_r <0$ in the neighbourhood of
$\lambdao=0.25$~mm. Finally, $\eps'_r >0$ but $\mu'_r <0$ around
$\lambdao=0.31$~mm. Of course, both $\eps''_r>0$ and $\mu''_r>0$
for all $\lambdao$.

Detailed calculations confirm that the spectral region wherein the
inequality (\ref{inequality}) is satisfied is larger than the
middle region (wherein both $\eps'_r <0$ and  $\mu'_r <0$). The
former cover parts of the adjoining regions, in which either
(\ref{inequalitye}) or (\ref{inequalitym}) holds.

In the five spectral regions identified, the isotropic
dielectric--magnetic medium would respond differently to
monochromatic electromagnetic excitation. Suppose that a plane
wave is normally incident on a half--space occupied by this
medium. The reflectance $R(\lambdao)$ is then given by the
standard expression
\begin{equation}
\label{Rl} R(\lambdao) =\Bigg\vert
\frac{+\sqrt{\mu_r(\lambdao)/\eps_r(\lambdao)}-1}
{+\sqrt{\mu_r(\lambdao)/\eps_r(\lambdao)}+1}\Bigg\vert^2\,,
\end{equation}
where $0\leq R\leq 1$ for all $\lambdao$ by virtue of the
principle of conservation of energy. The reflectance spectrum
calculated with the constitutive parameters used for Figures 2a
and 2b is shown in Figure 3. The reflectance is markedly high in
the two regions wherein $\eps'_r$ and $\mu'_r$ have opposite
signs, but not in the other three regions. The reflectance is
particularly low in the leftmost and the rightmost regions
($\eps'_r >0$ and  $\mu'_r >0$) because the ratio
$\mu_r(\lambdao)/\eps_r(\lambdao)$ is close to unity therein.
However, the reflectance is somewhat higher in the central region
($\eps'_r <0$ and  $\mu'_r <0$) because
$\vert\mu_r(\lambdao)/\eps_r(\lambdao)\vert < 0.25$.

\section{Conclusions}
In this pedagogical note, we have derived a general condition for
the phase velocity to be oppositely directed to the power flow in
isotropic dielectric--magnetic mediums in which the only
constraints on the values of the relative permittivity and
relative permeability are those imposed by causality. In this
regard, the topical case of mediums in which $\eps_r$ and $\mu_r$
have negative real parts is seen to be a sufficient, but not
necessary, condition for such propagation, as noted in the
comments succeeding (\ref{inequality}). An illustrative case study
has shown that there are, in general, five distinct spectral
regions, characterized by the various sign combinations of the
real parts of $\eps_r$ and $\mu_r$.

\section{Acknowledgement}
MWM  acknowledges the support of the {\em Engineering and Physical
Sciences Research Council\/} of Great Britain (EPSRC grant no.
GR/R55078/01).

\vspace{2cm}


\begin{center} {\bf Figure Captions} \end{center}

\vspace{1cm}

\noindent {\bf Figure 1}. Argand diagram parametrically displaying
${\mu}_{eff}(\omega)$ from equation (\ref{mu}), with $\Gamma =
0.1{\omega}_0$ and $F=0.5$. On taking (\ref{mu}) as a model
resonance form for meta--materials, the plot can also be regarded
as displaying the effective permittivity ${\eps}_{eff}$, for which
the resonance at $\omega = {\omega}_0$ is unlikely to coincide,
and hence the arrows indicating ${\eps}_{eff}$ and ${\mu}_{eff}$
do not coincide in general. The complex number
$n_+=+\sqrt{{\eps}_{eff}{\mu}_{eff}}$, while the corresponding
index $n_-=-\sqrt{{\eps}_{eff}{\mu}_{eff}}$ is not shown. The dots
indicate equi--spaced frequencies from $\omega = 0$ to $\omega =
2{\omega}_0$.
\bigskip

\noindent {\bf Figure 2}. (a) Real parts of the relative
permittivity and relative permeability according to  equations
(\ref{Leps}) and (\ref{Lmu}), respectively, when $p_e=1$,
$p_m=0.8$, $N_e = N_m=100$, $\lambda_e=0.3$~mm and
$\lambda_m=0.32$~mm. The significance of the identified five
regions of the spectrum is explained in the text. (b) Imaginary
parts of the relative permittivity and relative permeability
according to  equations (\ref{Leps}) and (\ref{Lmu}),
respectively, when $p_e=1$, $p_m=0.8$, $N_e = N_m=100$,
$\lambda_e=0.3$~mm and $\lambda_m=0.32$~mm.
\bigskip

\noindent {\bf Figure 3}. Plane wave reflectance $R(\lambdao)$
calculated with the constitutive parameters depicted in Figures 2.

\newpage
\begin{figure}[!ht]
\vskip 5cm
 \centering \psfull \epsfig{file=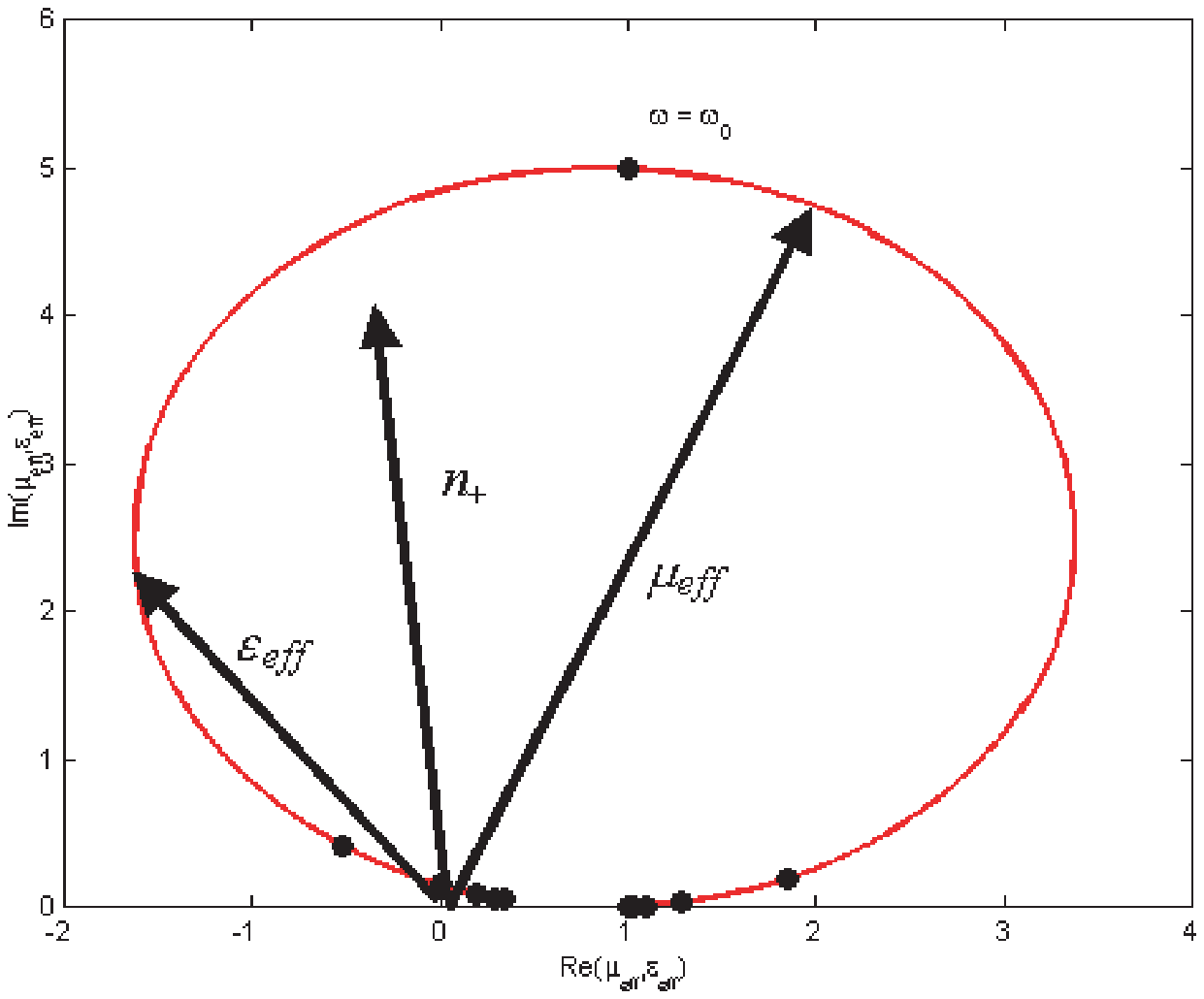,width=16cm }
\end{figure}

\newpage
\begin{figure}[!ht]
\centering \psfull \epsfig{file=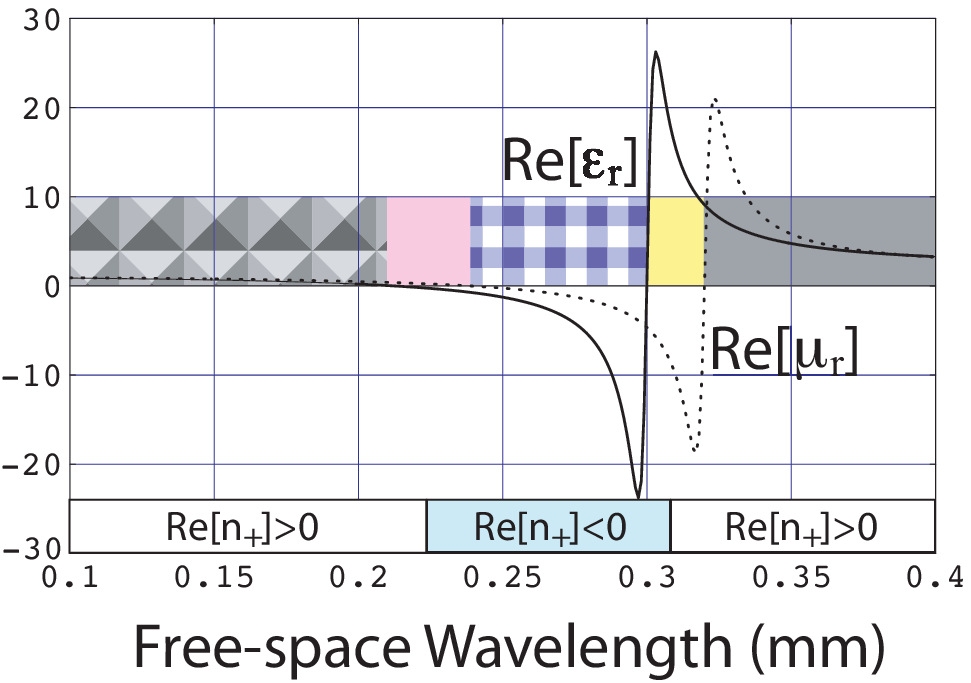,width=10cm } 
\vskip 2cm
\centering \psfull \epsfig{file=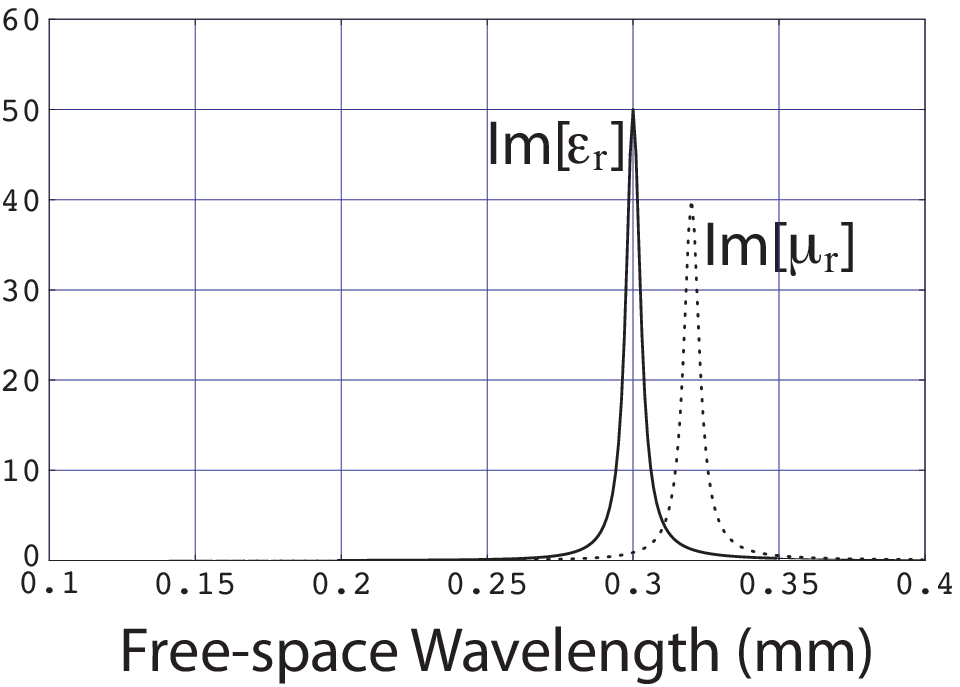,width=10cm }
\end{figure}

\newpage
\begin{figure}[!ht]
\vskip 5cm \centering \psfull \epsfig{file=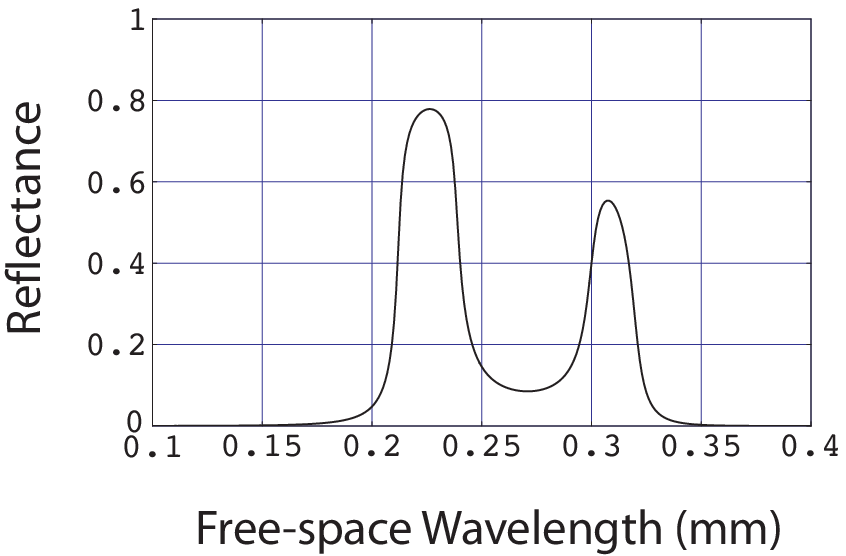,width=10cm }
\end{figure}


\section*{References}

\begin{thebibliography}{99}

\bibitem{Veselago}
Veselago V G 1968
{\em Soviet Physics Uspekhi\/} {\bf 10} 509--14

\bibitem{Schultz1}
Smith D R, Padilla W J, Vier D C, Nemat--Nasser S C and Schultz S
2000
{\em Phys. Rev. Lett.\/} {\bf 84} 4184--7

\bibitem{Pendry1}
Pendry J, Holden A J, Robbins D J and Stewart W J 1998
{\em J. Phys.: Condens. Matter.\/} {\bf 10} 4785--809

\bibitem{Pendry2}
Pendry J 1999
{\em Phys. Rev. Lett.\/} {\bf 85} 3966--9

\bibitem{Pendry5}
Pendry J, Holden A J, Robbins D J and Stewart W J 1999
{\em IEEE Trans. Microwave Theory Tech.\/} {\bf 47} 2075--84

\bibitem{Schultz2}
Shelby R A, Smith D R and Schultz S 2001
{\em Science\/} {\bf 292}(5514) 77--9

\bibitem{dewar}
Dewar G 2002 {\em Int. J. Modern Phys. B\/} {\bf 15} 3258--65

\bibitem{SPIE}
Lakhtakia A (ed) 1990 {\em Selected Papers on Natural Optical
Activity\/} (Bellingham, WA, USA: SPIE)

\bibitem{Smith}
Smith D R and Kroll N 2000 {\em Phys. Rev. Lett.\/} {\bf 85}
2933--6

\bibitem{Lindell}
Lindell I V, Tretyakov S A, Nikoskinen K I and Ilvonen S 2001
{\em Microwave and Opt. Tech. Lett.\/} {\bf 31} 129--33

\bibitem{Ziol}
Ziolkowski R W and Heyman E 2001 {\em Phys. Rev. E.\/} {\bf 64}
056625








\end{thebibliography}
\end{document}